\documentclass[final,5p,times,twocolumn]{elsarticle}
\usepackage{amssymb}
\usepackage{amsmath}
\journal{Journal of Magnetism and Magnetic Materials}

\begin{document}

\begin{frontmatter}

\title{Quantum effects and magnetism in the spatially distributed DNA molecules}

\ead{valentin.irkhin@imp.uran.ru}
\author[Ekb]{V.Yu. Irkhin}
\author[Urfu]{\fbox{V.N. Nikiforov}}

\address[Ekb]{M.N. Mikheev Institute of Metal Physics, Russian Academy of Sciences, 620108 Ekaterinburg, Russia}
\address[Urfu] {Moscow State University, Physics Department, Moscow 119992, Russia}

\begin{abstract}
Electronic and magnetic properties of DNA structures doped by simple and transition d- and f-metal ions (Gd, La, Cu, Zn, Au) are reviewed. Both one- and two dimensional systems are considered. A particular attention is paid to gadolinium and copper doped DNA systems, their unusual magnetism being treated. The problem of classical and quantum transport (including transfer of genetic information during replication and transcription) and electron localization in biological systems is discussed.
\end{abstract}


\end{frontmatter}

\section{Introduction}

The deoxyribonucleic acid (DNA)  has unique and sometimes mysterious properties which attract attention of not only biologists, but also physicists, chemists,  material scientists, nanotechnologists and engineers. The physical properties of natural DNA and metal-ion-inserted M-DNA have attracted much interest in recent years because of their scientific interest and potential practical applications. The bonding of DNA base pairs yields  the chemical foundation for genetics. Semiconducting or even metallic properties make long DNA chains suited for the construction of mesoscopic electronic devices and nanowires.

Last time, structural DNA nanotechnology has developed. Thus is an interdisciplinary field between engineering, physical, biological and medical sciences which becomes a useful tool for applications in spintronics and nanoelectronics. DNA and RNA (ribonucleic acid) provide a tractable and programmable system with control over intermolecular interactions, coupled with known structures for their complexes. Presently, construction of periodic and aperiodic lattices and topological structures is possible. DNA possesses predictable structures and extraordinary molecular recognition, can organize nanoparticles, biomacromolecules and nanomachines, The DNA nanostructures can be used in molecular and cellular biophysics, as biomimetic systems, in energy transfer, in diagnostics and therapeutics for human health (see, e.g., Refs. \cite{Seeman,Pinheiro}).


Besides possible practical applications, DNA provides a challenge concerning relations between modern condensed matter quantum physics and nature of living matter. The question about quantum effects in life was raised by E. Schroedinger \cite{Schroedinger} already in 1940s. He assumed that genetic mechanisms may be unexplainable  by classical  physics and discussed a possibility of quantum treatment based on discreteness and stability of quantum states.


At present, we have many popular communications and concepts, and also some serious publications concerning quantum effects in living systems (see the book \cite{Quantum} which contains both apologetic and critical material).
Quantum picture which includes life and consciousness is beautiful and appealing.
At the same time, there exist serious counterarguments from the scientific side. In particular, they are connected with the problem of relevant interaction, time and space scales,  and decoherence effects which destroy non-trivial quantum phenomena, including long-range entanglement (non-trivial means those beyond the standard atomic and molecular scales).
However, as noted by Zeilinger, ``It would be extremely non-trivial if quantum mechanics did not play a role in living systems, it would be the only area in which we know quantum mechanics is not at work'' (Plenary Debate, Ref.\cite{Quantum}, Part 5).

In the physical picture, the role of spin degrees of freedom is of considerable interest. This again leads us to quantum many-particle magnetism theory where occurrence, localization and interaction of the magnetic moments as related to electronic structure is one of central problems.
The important role of electron correlations in the DNA magnetism is discussed, e.g., in Ref. \cite{cor}.
In the present paper, we review the problem of electron transport and magnetism in pure and doped DNA systems. Some experimental data on Gd- and Cu-doped DNA will be also presented.

\section{Information transfer, conductivity and collective transport in DNA}


As mentioned above, size and timescales for various processes including biomolecules are of crucial importance for understanding life.
Treatment of this issue   is given in Ref.\cite{Quantum}).
Three different size scale ranges can be associated  with tissue-to-cell (1 cm-–-10 $\mu$m), cell-to-protein (10 $\mu$m–--10 nm) and protein-to-atom (10 nm–--0.1 nm) sizes.  Neurons can reach meters in length, so that simple classical diffusion along them may take many days.
DNA molecule, being stretched out, is about 2 meters long. However, it  is a rather subtle object, the scales for its main structures in compaction being: 2 nm (the DNA diameter), 10 nm (the nucleosome), 30 nm (the  chromatin fibre diameter), 10 $\mu$m (length of human chromosome  in metaphase).

As for timescales, they can be very different (in particular, after a rapid external influence, relaxation can be very slow).
Amino acid side chains oscillate in femtoseconds (10$^{-15}$ s), although longer time scale transitions may
last many seconds. Spontaneous  protein transitions
occur in the range from 10$^{-6}$ to 10$^{-11}$ seconds (microseconds to picoseconds).
There occurs  a clear separation of the scales for the relevant processes in water media: the times are $\tau_s \sim 100$ fs $\ll \tau_b \sim 10$ ps $\ll \tau_p \sim 10$ ns corresponding to and bulk solvent, bound water and proteins.
the corresponding spectral density for a typical chromophore demonstrates three distinct peaks (Ref.\cite{Quantum}, Part 2, Chapter 5).

One of the fundamental questions is how the information is preserved, transferred and repaired in DNA and protein structures. Among microscopic mechanisms, hydrogen bonds are often treated which constitute the pairing mechanism between DNA bases (see, e.g., \cite{Kornyshev}). Such mechanisms may include quantum tunneling. Besides classical approaches, a number of models of quantum information were discussed where DNA is treated as a chain of quantum bits (qubits); various qubit realizations in nucleotide systems as quantum superpositions of two possible base pair states are proposed (see \cite{Quantum,Pusuluk}).

Mayburov et al \cite{Mayburov} considered possible role of quantum effects in transfer of genetic information.
They assumed that the nucleotides selection during DNA replication is performed by means of proton tunneling between nucleotide and DNA-polimerase bound by hydrogen bonds. This mechanism is sensitive to the structure of nucleotide hydrogen bonds, so that only one nucleotide sort is captured by DNA-polimerase in each event.


Koruga \cite{Koruga} states that DNA and proteins could not be investigated separately: in  quantum  entanglement  approach  they  have to be considered as a unique system proteins are ``second side of DNA code''. Since the peptide plane is a  part  of  a  protein,
DNA-protein  coding (information) process can be treated as a synergy of classical and quantum effects.
Cancer is considered as violation of this synergy.
Basic element in this treatment is hydrogen bonds which  are  a considered as a
link  between  classical  and  quantum  behaviors on  the  molecular level  (carcinogens disallow the hydrogen bonds to be reformed).
A simple estimation of the action for  two coupling atoms in peptide plane, $h^* \sim 10^{-31}$ J s  \cite{Koruga}, in comparison with the Planck constant leads to the conclusion that both classical and quantum behavior is possible.

Genetic recombination in eukaryotes requires the pairing of homologous chromosomes to allow accurate molecular exchanges between chromosome pair, but the mechanism of the precise pairing  ``remains mysterious'' \cite{Falaschi}.
Baldwin et al \cite{Baldwin} observed spontaneous segregation
of the two kinds of DNA within each spherulite, so that nucleotide sequence recognition occurs between double helices separated by water in the absence of proteins at a large distance (about 2 nm). These authors proposed a mechanism based on electrostatic interactions: in-register alignment of phosphate strands with grooves on opposing DNA minimizes unfavorable electrostatic energy between the negatively charged phosphates and maximizes favorable interactions of phosphates with bound counterions. In other words, identical DNA double helixes have matching curves, so that they repel each other the least  (see also \cite{Kornyshev1} and review paper \cite{Kornyshev} which treats ``mesoscopic electrostatics'').
At the same time, other mechanisms, including quantum ones, can be considered (the situation is reminiscent of entanglement between DNA molecules).  Mazur \cite{Mazur} proposed recently that direct molecular recognition between homologous double stranded DNA is possible through the formation of short quadruplexes due to direct complementary hydrogen bonding of major-groove surfaces in parallel alignment.

Kurian et al \cite{Kurian} proposed that coherent vibrations in DNA and collective modes are significant for genomic and biological metabolism. Collective electronic behavior in the DNA helix generates coherent oscillations, quantized through boundary conditions imposed by the endonuclease, that provide the energy required to break two phosphodiester bonds. Such quanta may be preserved in the presence of thermal noise and electromagnetic interference through decoherence shielding. Clamping energy owing to the enzyme decoherence shield is comparable with the dipole-dipole oscillations in the DNA recognition sequence. This physical mechanism is assumed to explain long-range correlation between states in DNA sequence across spatial separations of any length which is a characteristic signature of quantum entanglement.

A model for the information transfer from DNA to protein using quantum information and computation techniques was proposed by Karafyllidis \cite{Karafyllidis}. On the DNA side, a 64-dimensional Hilbert space is used to describe the information stored in DNA triplets (codons). A Hamiltonian matrix is constructed for this space, using the 64 possible codons as base states. The eigenvalues of this matrix are not degenerate. The information on the protein side is also described by a 64-dimensional Hilbert space, but the eigenvalues of the corresponding Hamiltonian matrix are degenerate. Each amino acid is described by a Hilbert subspace.

From the physical point of view, the information transfer problem is related to conductivity of DNA molecules. Is it metal, semiconductor or insulator?  This question is also of importance for the problem of DNA sequencing (which uses electrical measurements) and for DNA based molecular electronics. For  natural DNA the issue remains contradictory \cite{Fink,Storm,Kasumov,Endres,Abdalla}.
In some investigations DNA was found to be insulating, even when the molecules had perfectly ordered base pairs.

Fink and Schoenenberger \cite{Fink} performed direct measurements of electrical current as a function of the potential applied across a few DNA molecules associated into single ropes at least 600 nm long and found good semiconducting properties (the resistivity values are comparable to those of conducting polymers).
Kasumov et al \cite{Kasumov} argue that interaction between molecules and substrate is a key parameter which determines the conducting or insulating behavior of DNA molecules. Strongly deformed (``dead'') DNA molecules deposited on a substrate, whose thickness is less than half the native thickness of the molecule, are insulating. At the same time, ``living'' molecules which keep their native thickness are conducting up to very low temperature, although demonstrating a non-ohmic behavior characteristic of a 1D conductor with repulsive electron--electron interactions.
Long-range transport in DNA molecules can be achieved through interaction with a disconnected array of metallic Ga nanoparticles, and even induced superconductivity could be observed \cite{Kasumov1}.

There is no theoretical basis for truly metallic DNA, but very small energy gaps might arise in the DNA-water-counterion complex, leading to thermally activated conduction at room temperature \cite{Endres}.
According to Ref. \cite{Abdalla}, localization of charge carriers inside potential wells in the lowest unoccupied and highest occupied molecular orbits (LUMO and HOMO) in DNA is described by several disorder parameters. Experimental data fitted with the presented model give evidence that the free carriers in the LUMO and HOMO are responsible to make the DNA molecule conductor, insulator or semiconductor. The localized charge carriers are characterized by four different types of relaxation phenomena with the corresponding thermal activation energies of 0.56, 0.33, 0.24, and 0.05 eV.

In his famous book ``What is life?'' \cite{Schroedinger} (which was written before Watson and Crick discovering the double helix genetic code) Schroedinger assumed that a molecular system carrying information should have regular structure but be aperiodic.
Indeed, each DNA  macromolecule is a long sequence of discrete site potentials —- of basic nucleotides (codons) ATCG (Adenine, Thymine, Cytosine, and Guanine). In the system of successive nucleotides, local levels with different depths are formed. DNA molecule is a stochastic sequence, the main feature of which being long-range correlation. Each base modulates a level inside the forbidden band. Ionization energies of the nucleotide are known: $\epsilon_A$ = 8.24, $\epsilon_T$ = 9.14, $\epsilon_C $= 8.87, and $\epsilon_G$ = 7.75 eV.

Tunneling along the chains depends appreciably on the level positions. The electron transport along this sequence occurs due to hopping between the neighboring nucleotides. This justifies the application of the tight-binding model.
The high conductivity in the case of large tunneling may result in a high rate of information transfer. Since most of mutations in DNA are successfully healed, one may assume the existence of charge transport through delocalized states that are responsible for the transfer of information at long distances. Then the localization length and the conductance of a given segment of the DNA molecule are directly related to the genetic information stored in this segment.

DNA conductivity in the  theoretical approach based on the electron localization in correlated disordered potentials was treated in Refs.\cite{Krokhin,Izrailev}.
The length of a genetic mutation is known to be relatively short (about 10 base pairs) as compared to the length of a gene (10$^3$--10$^6$ base pairs). Since most of the mutations in DNA are successfully healed, one may assume the existence of charge transport through delocalized states that are responsible for the transfer of information at long distances. According to Ref. \cite{Krokhin}, the exons (the parts where the genetic information is written) have narrow bands of extended (practically delocalized) states. At the same time, for the introns (the parts without apparent information for protein synthesis, the DNA ``dark matter'')  all the states are characterized by well-localized electron wave functions. Thus the localization length and the conductivity of a given segment of the DNA molecule are directly related to the genetic information stored in this segment.
For most of the energies the localization length inside the exon region exceeds by order of magnitude the localization length inside the intron region.
Thus DNA segments that apparently do not carry genetic code may not contain delocalized states in the energy spectrum, remaining insulators.

It should be noted that the role of introns in the genomic output is now reconsidered: it is supposed that non-proteincoding RNAs (ncRNAs) derived from introns of protein-coding genes and the introns and exons of non-protein-coding genes constitute the majority of the genomic programming in the higher organisms \cite{Mattick}.
It is believed sometimes \cite{Koruga} that the ``junk'' intron sequences (which are just characteristic for complex organisms \cite{Mattick}) may be active regulatory factor of system complexity trough microtubules (centrioles) and water in living systems (a concept, similar to that by Hameroff-Penrose who treated also consciousness problem \cite{Quantum,Penrose}).

In the simple band picture DNA has a valence band composed of nucleobases
HOMO (highest occupied molecular orbital) orbitals and a conduction band composed of nucleobases LUMO orbitals. They are separated by an energy gap of  4 eV causing the macromolecule to be an insulator. Therefore, in order to improve conducting properties of DNA one should add either additional electrons in the conducting band or additional holes in the valence band. A possibility of electron doping of DNA emerged with discovery of metal ion modified DNA (M-DNA),  a complex which DNA forms with transition metal cations \cite{Lee}.
A novel way of engineering DNA molecules involves substituting the imino proton of each base pair with a metal ion to obtain M-DNA with changed electronic properties.
Rakitin  et al \cite{Rakitin} reported a direct evidence of metallic-like conduction through 15 mm long Zn-DNA. On the other hand, the same measurement performed on native B-DNA (before conversion into M-DNA) showed semiconducting behavior with a band gap of a few hundred meV at room temperature.  However, these conclusions were criticized by Mizoguchi et al \cite{Mizoguchi}.

Measurement of the current-voltage characteristics of the Au-doped DNA molecules demonstrate that they exhibit a higher conductivity than undoped DNA \cite{Au}. The authors propose that Au doping will be a promising technique to control the conductivity of DNA molecules.
Further examples and details are considered in the next Section.

\section{Magnetism and transport in doped DNA systems}

A number of authors discuss unusual magnetic properties of natural DNA. According to Ref.\cite{magn}, they depend on the molecular structures: $\lambda$-DNA in the natural (wet) B state demonstrates paramagnetic behavior below 20 K which is nonlinear in an applied magnetic field, whereas in the A-DNA (dried) state it remains diamagnetic down to 2 K. The authors propose a mesoscopic orbital paramagnetism (that may be related to the proximity induced superconductivity) as an explanation and discuss its relation to the existence of long-range coherent transport in B-DNA at low temperatures.

Spin susceptibility $\chi_{\rm spin}$ and Lande g-factor value in Zn-doped DNA measured by Omerzu et al \cite{Omerzu} have a broad local maximum centered at 200 K and  are practically constant in the temperature interval between 100 and 300 K; $\chi_{\rm spin}$ does not follow a Curie-Weiss law expected for a system of localized spins. Below 100 K the g value starts to continuously increase with decreasing temperature suggesting a development of an internal local magnetic field.

Thus the Zn-DNA complex shows characteristics of a strongly correlated 1D electron system with antiferromagnetic spin correlations and high electron density in the conduction band \cite{Omerzu}. The 1D electron system is realized by transferring electrons to empty levels of DNA nucleobases. The electrons are not localized above 100 K and start to localize only at low temperatures below 50 K. Thereby the physics of strongly correlated systems becomes introduced into the field of doped DNA.

The situation was reconsidered, and some objections connected with the influence of magnetic impurities and occurrence of nonlinear paramagnetism in the dehydrated freeze dried (FD) Zn-DNA were put forward by Kumeta et al \cite{Kumeta}. The possible electronic states of the dehydrated FD Zn-DNA at room temperature are some nonmagnetic spin singlet ground states with a high transition temperature. At the same time,  the hydrated Zn-DNA was concluded to have a highly correlated $\pi$-band with a width of 0.24 eV. This order of the bandwidth is usual in organic systems, such as molecular conductors and fullerides. The change with hydration can be due to screening of the intersite Coulomb repulsion between the $\pi$-electrons of the neighboring base pairs by the water molecules.

$^{157}$Gd is a potential perspective agent for neutron capture cancer therapy (NCT, see Ref. \cite{Irkhin}). A higher gadolinium accumulation in cell nuclei as compared with cytoplasm takes place. This is significant for prospective NCT because the proximity of Gd to DNA increases the cell-killing potential of the short-range high-energy electrons emitted during the neutron capture reaction.  Linear DNA can bind a large number of Gd$^{3+}$ ions, because, in addition to phosphates, these ions can be inserted between the base pairs.

Gd- and La-doped DNA was investigated by Nikiforov et al \cite{Irkhin}. The measured magnetic susceptibility was fitted as $$\chi =\chi_0+C/(T-\theta).$$
For pure DNA, the signal from the sample has paramagnetic character ($\chi_0  > 0$), the Curie-Weiss contribution being small. The paramagnetism can be explained by the presence of the terminal OH groups with unsaturated oxygen bonds. The smaller (in particular with ``pure'' DNA) paramagnetic contribution to $\chi_0$ in La-doped DNA is likely owing to diamagnetic La$^{3+}$ ions. Besides that, La ions can interact with magnetic residuals (phosphate complexes, terminal residuals and so on) and ``heal'' the magnetic radicals. The same situation can occur for Gd ions where the dependence of $\chi_0$ on Gd concentration is non-monotonous. For large Gd concentration, estimated value of $\chi_0$ is  positive, which can be explained by prevalence of Van-Vleck contribution to magnetic susceptibility in comparison with the diamagnetic contribution.

The number of paramagnetic centers was calculated from the temperature dependence $\chi (T)$ in the range 77--300 K. The paramagnetic Curie temperature $\theta$ demonstrates finite values (Fig.1). Thus for Gd-DNA, the temperature dependence of magnetic susceptibility indicates the presence of interaction between Gd$^{3+}$ magnetic moments.

\begin{figure}[h]
\centering
\includegraphics[width=0.49\textwidth]{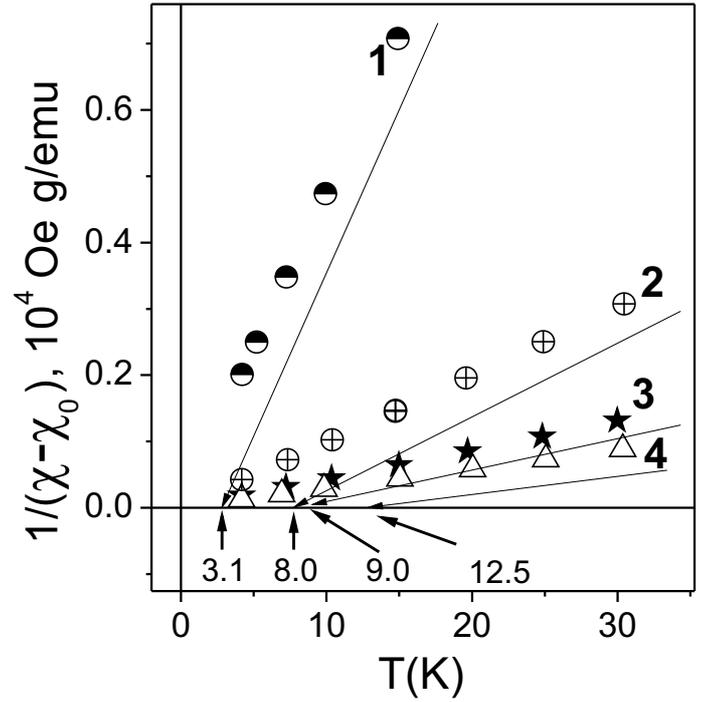}
\caption{Temperature dependence of magnetic susceptibility for four samples of Gd-doped DNA with different Gd concentrations \cite{Irkhin} at low temperatures. Lines 1--4 are Curie-Weiss extrapolations from high temperatures for the samples 1--4; arrows show the corresponding values of paramagnetic Curie temperature.
}
\label{fig:1}
\end{figure}

According to geometry of a DNA molecule (the step distance between protein bases is 3.4 \AA), the distance between atoms Gd on phosphates groups will be about 5--10 \AA. At the same time, appreciable values of $\theta$ indicate considerable interaction between Gd magnetic moments (no crystal field influence is expected for Gd$^{3+}$  ions). With further increase of Gd concentration (sample 5) $\theta$ can fall, which may mean destruction of conductivity in DNA chains. Thus the occurrence of finite $\theta$ is unlikely to be due to merging of close Gd atoms. Usual exchange interactions (like superexchange in insulators) between Gd ions can hardly work at the large separations. On the other hand, the RKKY interaction (characteristic for metals) can operate, being especially long-range in the one-dimensional case. Indeed, in this case the exchange parameter behaves at large distances $R$ as $$J_{\rm RKKY}(R) \sim \cos (2k_FR)/R$$
($k_F$ is the Fermi quasimomentum),
instead of $1/R^3$ and $1/R^2$ in the  three-dimensional and two-dimensional cases, respectively. Thus the RKKY interaction can occur for conducting DNA molecules. The interchain magnetic exchange interaction seems to be less significant, so that nearest Gd ion on other helix (second chain) will be chaotically located on rather long distances.

By analogy with the information transfer, we can assume that a long-range exchange between magnetic moments via current carriers may occur, although we cannot indicate precisely the mechanism of mediation of interaction (in particular, RKKY exchange, as for usual metals, or double exchange in a narrow conduction bands). Thus DNA chains with Gd ions located on the helix spatially ordered phosphate complexes seem yield an example of one-dimensional conductor with interacting magnetic moments.

A particular attention should be paid to magnetism of Cu-doped DNA systems \cite{Nik,Evd,Irkhin,Dugasani}. In particular, copper ions are nondestructive markers suitable for EPR control (in particular, of transport in cells). The incorporation of copper ion Cu$^{2+}$ into DNA allows the transfer of functionality, including thermal, electrical, magnetic, optical and chemical properties.

\begin{figure}[h]
\centering
\includegraphics[width=0.49\textwidth]{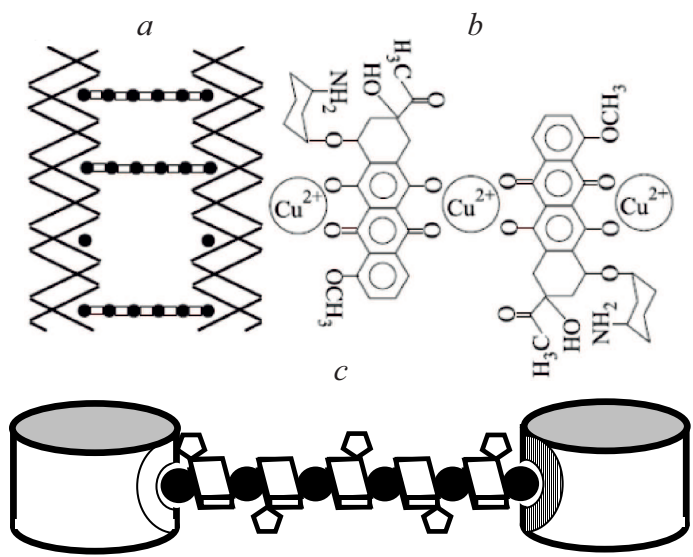}
\caption{
(a) Scheme of nanobridge between DNA molecules. (b) Chelate complex Cu$^{2+}$--DAU--Cu$^{2+}$ as central part of a nanobridge. (c) Top view of two DNA molecules and nanobridge between them. Each nanobridge is consisting of six Cu$^{2+}$ ions (dark spots) and five DAU molecules \cite{Nik,Evd}
}
\label{fig:2}
\end{figure}

It is known that DNA molecules form a cholesteric phase possessing high conductivity and self-organization \cite{Evd}. This phase may be the basis for the construction of organometallic complexes. The structure of the liquid crystal complexes is formed when introducing Cu$^{2+}$ ions into DNA. 
One quasinematic layer, formed by DNA molecules, can be treated as a set of simple equivalent matrices consisting of reaction centers. Then the formation of nanoconstructions can be considered as coordinated adsorption on reactive centers. This means that a nanobridge starting at the reactive center of one DNA molecule is terminated at the reactive center of the nearest NA molecule owing to spatial phasing of neighboring DNA molecules in the quasinematic layer. According to the X-ray study, DNA molecules are ordered in the particle at distances of 30--50\AA, i.e. acquire the properties of a crystal, molecules in the neighboring layers being mobile. Such a combination of properties allows this structure to be called as ``liquid-crystalline'' \cite{Evd}.

As a result of formation of nanobridges between the neighboring DNA molecules,
 nanoconstructions with fixed two- and three-dimensional spatial structures can occur  \cite{Nik,Evd}.
One can  suppose that all the copper ions are participating in the formation of nanobridges within nanoconstructions. Then each nanobridge, located in each helical turn between the neighboring DNA molecules, contains approximately six copper ions and five daunomycin (DAU) molecules (Fig. 2).

Appreciable value of $\theta$ about $-1$ K (which depends somewhat on the way of extracting the  contribution $\chi_0$)
indicates again existence of interaction between magnetic moments.
Thus the system demonstrates unusual magnetism which cannot be reduced to that of isolated magnetic ions: there occur exotic excitations characteristic of low-dimensional arrays. There is also an analogy with organic conductors and copper-oxide perovskites which are strongly correlated electronic systems.

In Ref. \cite{Tanaka}, one to five Cu$^{2+}$-mediated base pairs of hydroxypyridone nucleobases were incorporated into the middle of a double-stranded DNA, resulting in the formation of a magnetic chain by the lineup of mono- to pentanuclear Cu$^{2+}$ complexes stacked within the DNA helix. The Cu$^{2+}$  ions in each complex were coupled ferromagnetically via unpaired d-electrons.

Dugasani et al \cite{Dugasani} constructed copper ion modified DNA thin films as fully covered DNA polycrystalline structures with a divalent Cu on glass substrates and studied their magnetic properties at temperatures down to 8 K. DNA films with 5 millimolar (mM) Cu$^{2+}$ optimum concentration exhibited a strong ferromagnetic nature, the hysteresis loop being observed. The magnetization curves at higher temperatures region indicated ferromagnetism with Curie temperature above 300 K. The SQUID measurements showed a tendency for a strong ferromagnetic interaction between the neighboring Cu ions.

The ferromagnetic behavior could be attributed to the presence of small magnetic dipoles of Cu$^{2+}$ located inside the DNA bases, which interact with their nearest neighbors inside the DNA molecules. With the increase of the Cu$^{2+}$ concentration, magnetization and $\chi$ decreased.  Possibly, an increased number of Cu$^{2+}$ occupying adjacent positions results in antiferromagnetic alignment.

\section{Conclusions}

 The physics of biological molecules turns out to be very complicated and includes many puzzling phenomena \cite{Kornyshev}.
Although a number of fundamental and technical problems remain unsolved, perspectives of the use of DNA based systems seem to be promising.
For example, artificially designed DNA nanostructures with various metallic ions can become a new class of material for electronic and magnetic devices (e.g., molecular magnets) in the near future \cite{Dugasani}.
Different magnetic effects (e.g., magnetoresistance) can be used also for biosensing \cite{Baselt,Miller}. DNA can be present as a part of molecular recognition process. The understanding of the mechanisms of molecular binding of the analyte DNA and the magnetic markers is crucial for good selectivity and sensitivity which cannot be reached without understanding of these mechanisms \cite{Schrittwieser}.

At present, a possibility of creating new room-temperature ferromagnetic organic materials technologies based on DNA science is discussed \cite{book}.

Peculiar feature of  low-dimensional systems is a tendency towards localization of electronic states even for weak disorder and unusual magnetism with strong short-range order. The DNA magnetic system (in particular, in M-DNA) has high organization and demonstrates both local-moment and itinerant-electron features.
DNA conductivity, which is varied depending on organization of the codon sequences, gives a possibility to create modern fast and stable nonvolatile memory devices on the basis of localized codon levels.

The low-dimensional systems under consideration  can be treated as strongly correlated ones with exotic properties and possess unusual excitation spectrum. Modern many-particle concepts of condensed matter physics include topological phases, strings and tensor networks, the matter being considered as assembly of qubits -- simplest elements of quantum information \cite{Wen,Wen1}. Such systems demonstrate quantum entanglement and long-range correlations. These issues may concern mechanisms of information transfer and storage in biological systems, and also of short-term and long-term memory of living beings.


The authors are grateful to B.L. Oksengendler, Yu.N. Shvachko, Yu.N. Skryabin, V.A. Lange and G.V. Kurlyandskaya for valuable discussions.
The research was carried out within the state assignment of FASO of Russia
(theme ``Quantum'' No. 01201463332).

\end{document}